\documentclass[english,prd,twocolumn,superscriptaddress,nofootinbib,preprintnumbers]{revtex4}
\usepackage[latin1]{inputenc}
\usepackage{graphicx}
\usepackage{amssymb}
\usepackage{color}

\usepackage{amsmath}
\usepackage{amsfonts}
\usepackage{dcolumn}
\usepackage{hyperref}

\def\be{\begin{equation}}
\def\ee{\end{equation}}
\def\bea{\begin{eqnarray}}

\def\eea{\end{eqnarray}}
\def\beq{\begin{eqnarray}}
\def\eeq{\end{eqnarray}}

\begin{document}

\title{Generic f(R) theories  and classicality of their scalarons }

\author{Radouane Gannouji}
\affiliation{Department of Physics, Faculty of Science, Tokyo
University of Science, 1-3, Kagurazaka, Shinjuku-ku, Tokyo 162-8601,
Japan}

\author{M. Sami\footnote{On leave from Centre for theoretical
physics, Jamia Millia islamia, New Delhi}}
\affiliation{Kobayashi-Maskawa Institute for the Origin of Particles
and the Universe, \\Nagoya University, Nagoya 464-8602, Japan}

\author{I. Thongkool}
\affiliation{Harish-Chandra Research Institute, Chhatnag Road,
Jhusi, Allahabad-211019, India}

\begin{abstract}
In this Letter, we study quantum stability bound on the mass of
scalaron in generic theories of $f(R)$ gravity. We show that in
these scenarios, the scalaron mass increases faster with local
density of the environment than one-loop quantum correction to it
thereby leading to violation of quantum bound on the chameleon mass.
The introduction of quadratic curvature corrections in the action
are shown to stabilize the model.

\end{abstract}

\maketitle

\section{Introduction}

The late time cosmic acceleration \cite{Riess:1998cb,Perlmutter:1998np} has recently been accepted as one
of the fundamental phenomena of nature
 whose underlying cause
remains yet to be unfolded. The standard lore preaches that the late
time acceleration  is caused by the presence of a cosmic fluid with
large negative pressure; the cosmological
 constant $\Lambda$ \cite{Martin:2012bt} presents a distinguished example of such a fluid.

As an alternative to cosmological constant, a variety of scalar
field models were
 investigated with a hope
 to alleviate the fine tuning  and coincidence problems associated
 with the model without assigning a fundamental reason to switch off $\Lambda$.
 Unfortunately, the scalar field dark energy models are not
entirely problem free, assumptions
  about model parameters/tuning are tacitly made in these models.\\

There is an alternative  school of thought in cosmology which
advocates the need for paradigm shift and believes that cosmic
acceleration results from large scale modification of gravity \cite{Clifton:2011jh}. Such
a proposal sounds healthy as general theory of relativity, which
passes the solar test with great precision, is often extrapolated to
large scales where it was never verified directly.

 One
of the schemes of large scale modification based upon
phenomenological considerations is provided by $f(R)$ theories of
gravity \cite{Sotiriou:2008rp,DeFelice:2010aj}. These theories
essentially contains an additional scalar degree of freedom apart
from graviton. Indeed, $f(R)$ theories are conformally equivalent to
Einstein theory plus a canonical scalar degree of freedom dubbed
scalaron whose potential is uniquely constructed from Ricci scalar.

It is interesting to note that $f(R)$ gravity is Ostrogradski ghost
free despite the equations of motion being of fourth order as there
are
enough number of constraints to protect the theory.\\

However, it should further be ensured that the graviton and the
scalaron are well behaved which imposes restrictions on the
functional form of $f(R)$. Namely,
  the generic $f(R)$ theories should
satisfy, $f'(R),f''(R)>0$ in order to avoid the ghost and tachyonic
modes. Secondly, these theories  should reduce to $\Lambda CDM$ in
the high density regime in order to comply with the local gravity
constraints. The class of models proposed by Hu-Sawicki and
Starobinsky (HSS) \cite{Hu:2007nk,Starobinsky:2007hu}(see also
Ref.\cite{appleby} on the same theme) satisfy the said requirements
and are of great interest in $f(R)$ theories.

In this scenario, the scalar degree of freedom is non-minimally
coupled to matter in Einstein frame and hence it might conflict with
the local physics which does not see a fifth force. Thus if all is
to be well, the scalaron should acquire a heavy mass in local
environment in order to suppress the fifth force and become light
with mass of the order Hubble constant today to be relevant to late
time cosmic acceleration which means that the scenario asks for a
chameleon.

In $f(R)$ theories, the scalaron mass naturally acquires density
dependence thereby allowing us to naturally implement the chameleon
mechanism by appropriately choosing the form of $f(R)$ giving rise
to higher values of scalaron mass for larger values of density of
the environment. The chameleon scenario, despite of its
attractiveness, is plugged with several difficult problems: The
scalaron mass might exceed the Planck mass by several orders of
magnitudes in high density configurations such as neutron stars, the
curvature singularity is easily accessible in the scenario and
requires ugly fine tuning for its cure\cite{probfr}. Being inspired
by Starobinsky's  original proposal \cite{staro}, the HSS model was
extended by adding quadratic curvature correction \cite{ext} to
address the said problems.\\

 The quadratic correction provides in a sense quantum correction
 to gravity sector which turns out to be important in the scenario
 under consideration. It becomes equally important to investigate
 whether the quantum 1-loop correction to scalaron potential
 remains  small as density of the environment increases.\\

 In this Letter we shall study the quantum stability bound
   for scalaron in Starobinsky $f(R)$ gravity model. We also
   address  the same issue in the framework of an extended scenario by incorporating the quadratic curvature
   corrections in the Starobinsky model.

\subsection*{Chameleon field}

Let us consider the following action in the Einstein frame

\begin{align}
\label{eq:action}
\mathcal{S}=\int d^4x\sqrt{-g}\Bigl[\frac{M_{\text{pl}}^2}{2}R-\frac{1}{2}(\nabla\phi)^2-V(\phi)\Bigr]\nonumber\\
+\mathcal{S}_m\Bigl[A(\phi)^2g_{\mu\nu},\Psi_m\Bigr]
\end{align}

The equation for the field $\phi$  which follows from the action can
be written as

\begin{align}
\label{eq:cons}
\Box \phi=\frac{dV}{d\phi}+(\rho-3P)A^3\frac{dA}{d\phi}
\end{align}

where $(\rho,P)$ are energy density and the pressure in the Jordan
frame. We consider this frame as our physical frame in which the
stress-energy tensor is conserved hence we assume that our
quantities are independent of the scalar field $\phi$. We note that
in the original paper \cite{Khoury:2003aq}, the authors defined a
conserved density in the Einstein frame for a FLRW space-time. The
definition that we adopt here gives a definition of the effective
potential for any background (also in presence of pressure) and
within this definition the effective mass of the chameleon field is
the mass of the scalaron in $f(R)$. It is however clear that because
in most of the cases $A\simeq 1$ the quantities in the two frames
are very close.

The eq.(\ref{eq:cons}) can be cast in the form

\begin{align}
\Box \phi=\frac{dV_{eff}}{d\phi}
\end{align}

where $V_{eff}=V+\frac{\rho-3P}{4}A^4$.

The existence of the chameleon mechanism depends on the form of the
effective potential which in turn depends on the local density and
pressure. When pressure is negligible and density is large, the
scalar field may acquire a large mass for a suitably chosen
potential leading to suppression of the fifth force locally. The
scalaron mass is defined as usual

\begin{align}
m^2_{eff}=\frac{d^2V_{eff}}{d\phi^2}
\end{align}

The scalar field is assumed to be settled  in  the
minimum of the effective potential. It is therefore simple to recast
the effective mass in the following form

\begin{align}
m^2_{eff}=V''-V'\Bigl(3\frac{A'}{A}+\frac{A''}{A'}\Bigr)
\end{align}
To avoid a ghost instability, we require that
$V''/V'>3\frac{A'}{A}+\frac{A''}{A'}$.

In the simple case when $A$ is given by,  $A=e^{\beta\phi/M_p}$, we
have

\begin{align}
\label{eq:m}
m^2_{eff}=V''-4\frac{\beta}{M_p}V'
\end{align}
In what follows we shall consider the chameleon mechanism in $f(R)$
theories of gravity where it occurs naturally.

\subsection{Chameleon theory and $f(R)$ gravity}
\noindent

Let considering  $f(R)$ action in the Jordan frame,
\begin{equation}
\mathcal{S} =\frac{M_{pl}^2}{2} \int d^4x \sqrt{-g} f(R) + \mathcal{S}_{m} \lbrack g_{\mu\nu},\Psi_i \rbrack.
\end{equation}
We next use a conformal transformation

\begin{align}
\label{eq:conf}
g_{\mu\nu} \rightarrow f_{,R}~g_{\mu\nu} = e^{-2\beta\phi / M_{pl}} g_{\mu\nu}
\end{align}

with $\beta = -1/\sqrt{6}$, to transform  the action to the Einstein
frame

\begin{align}
\mathcal{S}_E = \int d^4x \sqrt{-g} \Bigl[ \frac{M_{pl}^2}{2} R - \frac{1}{2} \Bigl(\nabla\phi\Bigr)^2 -V(\phi) \Bigr]\nonumber\\
+ \mathcal{S}_{m} \Bigl[e^{2\beta\phi/M_{pl}}g_{\mu\nu},\Psi_i \Bigr],
\end{align}

where
\begin{equation}
\label{eq:vphi}
V(\phi) = M_{pl}^2\frac{R f_{,R} - f}{2f_{,R}^2}.
\end{equation}

The first and second derivatives of the potential $V(\phi)$ are given by

\begin{align}
\label{eq:Vp}
V,_\phi &= \frac{M_{pl}}{\sqrt{6}}\frac{2f-Rf_{,R}}{f_{,R}^2},\\
\label{eq:Vpp}
V,_{\phi\phi} &= \frac{1}{3f_{,RR}} \left ( 1 + \frac{R f_{,RR}}{f_{,R}}-\frac{4f f_{,RR}}{{f_{,R}}^2} \right ).
\end{align}

One can see that effective potential belongs to Chameleon theory as
($\phi$ directly couples to matter),
\begin{equation}
\label{eq:veff}
 V_{eff} = V(\phi) + \frac{\rho-3P}{4}A^4,
\end{equation}
where $\rho$ and $P$ are respectively the density and the pressure in Jordan frame and $A=1/\sqrt{f_{,R}}$.

It is easy to find that the minimum of the effective potential from
eq.(\ref{eq:Vp},\ref{eq:veff}),

\begin{align}
\label{eq:min}
2f-Rf_{,R}=\frac{\rho-3P}{M_p^2}
\end{align}

It is interesting to notice that the minimum of the potential is
invariant under the addition to the action of a $R^2$-term. We shall
use this aspect in the discussion to follow.

Also one can rewrite the effective mass (\ref{eq:m}) with the help of (\ref{eq:Vp},\ref{eq:Vpp}) as

\begin{align}
\label{eq:mass1}
m_{eff}^2=\frac{1}{3f_{,RR}}\Bigl(1-\frac{Rf_{,RR}}{f_{,R}}\Bigr)
\end{align}

which corresponds to the mass of the scalaron in the Einstein frame.
In fact in the Jordan frame, we have
$M^2=\frac{1}{3f_{,RR}}\Bigl(f_{,R}-Rf_{,RR}\Bigr)$ and therefore
$M=\sqrt{f_{,R}}m_{eff}=A^{-1}m_{eff}$ which is the standard factor
which relates the mass of the field  in Jordan to its counter part
in Einstein frame. We should emphasis  that we recover the mass of
the scalaron because we consider the effective potential instead of
the potential and also because we consider the density $\rho$ and
the pressure in the Jordan frame as defined in (\ref{eq:veff}).

We often encounter local densities much larger than cosmological
density $\rho_{cr}$ such as $\rho_{lab}\simeq 10 $ g$/$cm$^3$ and use
the classical description for scalar degree of freedom in $f(R)$
which assumes the quantum correction to scalaron potential to be
small. Following Ref.\cite{Upadhye:2012vh}, we shall now address the
issue of quantum stability in generic theories of $f(R)$ gravity.

\subsection*{Quantum stability bound}

The effective potential defined in (\ref{eq:veff}) depends on the
energy density and therefore on the position of space-time. At the
equilibrium, the field minimizes the potential $V_{eff,\phi}=0$,
hence the chameleon appears as a massive field ($V_{eff}\simeq
m_{eff}^2\phi^2/2$).{In Einstein frame, we have General Relativity
with a scalar field. In scenario under consideration, the curvature
scalar is small as seen later and the effects of the expansion are
negligible. Hence the model is close to  quantum field theory in
flat space-time where we can neglect the effects of gravity
\footnote{The gravity sector is the massless spin two particle
without the scalaron.} and quantize  the scalar field sector in
standard way. Also we should emphasize that we can always work in
the Einstein frame, as long as the conformal transformation is not
singular. Indeed this the case for the model studied  as  $R \simeq
\rho/M_{pl}^2$, to be demonstrated shortly.}

In \cite{Upadhye:2012vh}, the authors considered the one-loop
Coleman-Weinberg correction to the chameleon field potential.

\begin{equation}
\label{qb}
 \Delta V_{1-loop} (\phi) = \frac{m_{eff}^4 (\phi)}{64
\pi^2} \ln \left ( \frac{m_{eff}^2(\phi)}{\mu_0^2} \right ),
\end{equation}

where $\mu_0$ is a cut off mass scale. It can be chosen equal to the
mass of the field for a particular environment (density) which would
kill the quantum correction but the correction would reappear at
other values of the density.

At large values of density of interest to us or large mass of the
field, we can set log term to unity
\begin{equation}
\Delta V_{1-loop} (\phi) \simeq \frac{m_{eff}^4}{64\pi^2}.
\end{equation}

Since we expect quantum corrections to be small, we should have
\cite{Upadhye:2012vh} a small modification of the shape of the
potential $V$ which implies $\Delta V_{1-loop,\phi}/V_{,\phi}$ and
$\Delta V_{1-loop,\phi\phi}/V_{,\phi\phi}$  should be small.

Secondly, at the minimum of the effective potential
(\ref{eq:veff}), we have{\footnote{For simplicity we neglect the pressure}}

\begin{align}
V'(\phi)+\frac{\beta}{M_{pl}}\rho e^{4\beta\phi/M_{pl}}=0
\end{align}

from which we obtain

\begin{align}
\frac{d\phi}{d\rho}=-\frac{\beta}{M_{pl} m_{eff}^{2}(\phi)} e^{4\beta\phi/M{_{pl}}}
\end{align}

which gives

\begin{align}
\frac{\Delta V_{1-loop,\phi}}{V_{,\phi}}&\simeq \frac{M_{pl}^2}{96\pi^2\beta^2}\frac{1}{\rho}\frac{d m_{eff}^6}{d\rho}<1\\
\frac{\Delta V_{1-loop,\phi\phi}}{V_{,\phi\phi}}&\simeq \frac{M_{pl}^2}{96\pi^2\beta^2}\frac{d^2m_{eff}^6}{d\rho^2}<1
\end{align}

and after integration

\begin{align}
m_{eff} < \left ( \frac{48 \pi^2 \beta^2 \rho^2}{M_{pl}^2} \right )^{1/6} = 0.0073 \left (\frac{|\beta|  \rho}{10 \textrm{g cm}^{-3}} \right )^{1/3} \textrm{eV}.
\end{align}

where, the constant of integration is fixed to zero; we can demand
that the correction is zero for very low densities where the
effective mass is zero.
{ Let us briefly comment on the viability of Coleman-Weinberg one
loop correction used here for chameleon. It corresponds to the
quantum mechanically corrected  potential

\begin{equation}
V_{tot}=V_{eff}(\phi)+\frac{i}{2}\ln
det\big[\partial^2+m_{eff}^2)\big]
\end{equation}

where the first term represents classical part of the potential. The
quantum correction is formally divergent and requires ultraviolet
cut off. In case we use non-covariant  scheme of regularization,
say, Pauli-Willars regularization, with cut off $M_{uv}$, the
quantum correction is represented by three terms: (1)$ M^4_{uv}$,
(2) $m_{eff}^2M_{uv}^2$ and (3) $m_{eff}^4\ln( m_{eff}^2)$. The
first term can be absorbed in the definition of renormalization of
cosmological constant, the third term represents the one-loop
quantum correction to be used in the discussion to follow. However,
the second term is much larger than the second and would invalidate
usage of the quantum
bound based upon the third term only.\\

It is well known that the term quadratic in cut off is specific to
any regularization scheme which breaks Lorentz symmetry. In case of
gauge theories, the regularization scheme which does not respect the
underlying symmetry of the theory leads to wrong results
\cite{Rosenberg:1962pp}. Indeed, in the present context, the
regularized value of the quantum correction using dimensional
regularization gives rise to expression (\ref{qb}) without the
dangerous term quadratic in cut off. It is interesting to note that
we see similar features when we regularize the vacuum energy. In
fact the correction can be understood as the zero-point energy
density of the scalaron. It appears as a massive Klein-Gordon field
and gives for the energy density of the vacuum

\begin{align}
\rho=\frac{1}{2(2\pi)^3}\int d^3 k \sqrt{k^2+m_{eff}^2}
\end{align}

As we previously said, a regularization that do not respect the symmetries of the problem is incorrect \cite{Akhmedov:2002ts}, and produce the terms detailed beforehand. Hence a dimensional regularization of the energy density gives in the $\overline{\text{MS}}$ scheme

\begin{align}
\rho&=\lim_{d\rightarrow 4}~ \frac{\mu_0^{4-d}}{2 (2\pi)^{d-1}}\int d^{d-1}k \sqrt{k^2+m_{eff}^2}\nonumber\\
&\simeq \frac{m_{eff}^4}{64\pi^2} \ln\Bigl(\frac{m_{eff}^2}{\mu_0^2}\Bigr)+\cdots
\end{align}

where $\mu_0$ is introduced to clean up the units.\\

}

{ We shall hereafter would specialize to $f(R)$ gravity. We should
emphasis that the scalaron potential in general is a complicated one
and certainly does not belong to the class of renormalizable theory.
However, in the neighborhood of its minimum, the latter can be
approximated by a polynomial. Thus we can apply the quantum bound
obtained using the Coleman-Weinberg formula for effective
potential.}

 In $f(R)$, $\beta = -1/\sqrt{6}$, which implies that

\begin{align}
\label{mm}
 m_{eff} < 5.4~10^{-3} \left (\frac{\rho}{10 \textrm{g
cm}^{-3}} \right )^{1/3} \textrm{eV}.
\end{align}

Eq.(\ref{mm})  provides an upper bound on the mass of the field. As
we mentioned before, we shall be interested in the scrutiny of
generic $f(R)$ theories, namely HSS and would specialize to
Starobinsky parametrization for convenience.
\subsection*{ Parameters of Starobinsky model}

We are interested to study the quantum stability of Starobinsky
$f(R)$ gravity \cite{Starobinsky:2007hu}

\begin{equation}
\label{eq:star}
f(R) = R - \mu R_c \left \lbrack 1 - \left ( 1 + R^2/R_c^2 \right )^{-n} \right \rbrack.
\end{equation}

During the de-Sitter phase, the solution is described by (\ref{eq:min}) in an empty Universe.
The curvature scalar ($R_{dS}$) is solution of

\begin{align}
\label{eq:mu}
\mu = \frac{1}{2} \frac{x (1+x^2)^{n+1}}{(1+x^2)^{n+1}-1-(1+n)x^2},
\end{align}

where $x=R_{dS}/R_c$. Considering $R_c$ of the order the curvature scalar today we have $\mu\simeq 1$.

In the region of high density $(R \gg R_c)$, we have
\begin{equation}
\label{eq:starhigh}
f(R) \simeq R - \mu R_c \left \lbrack 1 - (R/R_c)^{-2n} \right \rbrack.
\end{equation}

It can easily be noticed from (\ref{eq:min}) that the minimum is
$R\simeq \rho/M_{pl}^2$ as in General Relativity.{Hence the
gravitational sector is exactly equivalent to the standard frame
work of General Relativity. The scalaron which settles at the
minimum of the effective potential has small variation around this
point because of the space dependence of the density of matter
(\ref{eq:min}). }

This translates to the chameleon field via its definition
(\ref{eq:conf}) and gives the minimum of the effective potential

\begin{align}
\label{eq:Rmin}
\frac{\phi}{M_{pl}}=\sqrt{\frac{3}{2}}\ln~f' \simeq \sqrt{\frac{3}{2}}\Bigl[f'(\frac{\rho}{M_{pl}^2})-1\Bigr]
\end{align}

Let us now consider the experimental bound that comes from the solar
system tests of the equivalence principle (LLR). Using the
thin-shell parameter \cite{Khoury:2003aq} for the Earth $\epsilon_{\textrm{th}}$ we have

\begin{equation}
\epsilon_{\textrm{th}} \equiv \frac{\phi_\infty - \phi_\oplus}{6|\beta| M_{pl}\Phi_\oplus} < \frac{8.8 \times 10^{-7}}{|\beta|},
\end{equation}

where $(\phi_\infty,\phi_\oplus)$ are respectively the minimum of the effective potential at infinity and inside the planet and $\Phi_\oplus$ the Newton potential for the Earth.

Using the value $\Phi_\oplus\simeq 7\times 10^{-10}$, the previous
bound translates into $\phi_\infty/M_{pl}<10^{-15}$, which after
using Eq.(\ref{eq:Rmin}) leads to

\begin{align}
\Bigl|f'(\frac{\rho_\infty}{M_{pl}^2})-1\Bigr|<10^{-15}
\end{align}

For the HSS model and with the density
 $\rho_\infty\simeq 10^{-24}$ g cm$^{-3}$
 and $R_c\simeq H_0^2$, we have $10^{-5 (2n+1)}<10^{-15}$ tells us that $n>1$ \cite{Tsujikawa:2007xu}.

We will show that for this set of parameters the quantum stability
is violated  in the Starobinsky model  as the mass of scalaron in
the model grows  fast with density and can easily cross the quantum
bound.

\subsection*{Quantum stability of f(R) gravity}

According to \cite{Upadhye:2012vh}, the bound on $m_{eff}$ obtained
using the 1-loop Coleman-Weinberg correction is given by

\begin{align}
\label{classicality}
m_{eff}(\rho) < 5.4\times 10^{-3} \left ( \frac{\rho}{10 \textrm{ g cm}^{-3}} \right )^{1/3}  \textrm{eV},
\end{align}

Also from Eq.(\ref{eq:mass1}), we can express the scalar field mass
in Starobinsky model as a function of the density $\rho$

\begin{align}
\label{eq:mass}
m_{eff}(\rho) \simeq \frac{1}{\sqrt{6\mu n(2n+1)}} \sqrt{R_c} \left ( \frac{\rho}{M_{pl}^2R_c} \right )^{n+1}.
\end{align}

where we assumed that the density is large enough compared to the cosmological density $M_{pl}^2R_c \simeq \rho_c \simeq 10^{-29}$ g cm$^{-3}$

From the previous discussion, we know that $\mu\simeq 1$ and $n>1$,
which gives

\begin{equation}
m_{eff}(\rho) \simeq 3\times 10^{-34} \left ( \frac{\rho}{\rho_c} \right )^{n+1} \textrm{eV}.
\end{equation}

At the cosmological density, $\rho \sim \rho_c\sim
10^{-29}$g/cm${}^3$, the quantum stability bound, $5 \times
10^{-13}$ eV, is larger than the scalar field mass $m_{eff} =
3\times 10^{-34}$ eV. However, the quantum bound $\propto
\rho^{1/3}$ while $m_{eff} \propto \rho^{n+1}$ (with $n>1$).  It is
therefore clear that the $m_{eff}$ will be excluded easily by this quantum stability bound at some high density.\\
Indeed, the scalar mass $m_{eff}$ is quickly excluded by the quantum
stability bound at  $\rho \approx 10^{-87 n/(2+3n)}$ g/cm${}^3$.

That means according to this bound, $f(R)$ gravity in the point of
view of chameleon theory is excluded in any typical dense medium,
i.e., in the air ($\rho_{_{\textrm{air}}} \sim 10^{-3}$ g/cm${}^3$).

\subsection*{ Extended Starobinsky model: Adding $\alpha R^2$ term}

In high density regime the quantum curvature corrections to
Einstein-Hilbert action become important. These corrections might
provide a cut off  to the scalaron mass. In what follows, we shall
consider the extension of Starobinsky model by  adding $\alpha R^2$
term to its action,

\begin{equation}
f(R) = R - \mu R_c \left \lbrack 1 - \left ( \frac{R}{R_c} \right )^{-2n} \right \rbrack + \alpha R^2,
\end{equation}

{This correction was briefly suggested  in the original
paper\cite{Starobinsky:2007hu}
  to avoid the problem of scalaron mass from
 becoming too large and being inspired by the Starobinsky's original idea, it was
 introduced in \cite{ext} as
 a solution to the Frolov singularity problem(see also Refs.\cite{ext1} on the same theme).}
As noticed above, the addition of this term does not change the
position of the minimum of the field but it will provide a natural
upper bound to the mass of the chameleon field.
 {Thus the gravitational sector is unchanged compared to
 general Relativity, we have the same curvature scalar but the scalar
 sector is modified because of this additional term. The shape of the effective potential is changed.}

In the regime of large densities, we have

\begin{align}
m_{eff}^2\simeq\frac{1}{6\alpha(1+2\alpha\rho/M_p^2)}
\end{align}

We should emphasize that the scalaron is massless in the Einstein
frame when the density diverge contrary to the Jordan frame where
the effective mass goes to $1/6\alpha$, this is because of the
conformal factor $f_{,R}$ and it will have no effect on the
following discussion. In fact, we consider hereafter $\rho \simeq \rho_{\textrm{lab}}$ which
gives in both frames $m_{\textrm{eff}}^2 \simeq 1/6\alpha$.

The classicality condition (\ref{classicality}) gives a lower bound
$\alpha\ge 6\times 10^3$ eV$^{-2}$ when according to the bound from
Big Bang  nucleosynthesis and CMB physics $\alpha \ll 10^{35}$ eV$^{-2}$ \cite{Zhang:2007ne}.
But the tightest bound comes from the E\"ot-Wash experiments which
implies that \cite{Kapner:2006si} $\alpha<4 \times 10^4$ eV$^{-2}$.
We still have a range of viability of the model as soon as we add
the quadratic curvature correction term in the action.

Also it should be noticed that $R^2$-term is introduced here as a
cure of late time cosmic dynamics. However, we know that $R^2$ can
gives rise to inflation at early epochs. And if the model is used to
describe an early acceleration phase we would have
\cite{Starobinsky:1983zz,Starobinsky:2007hu} $\alpha \simeq 10^{-45}
\Bigl(N/50\Bigr)^2$ eV$^{-2}$ where $N$ is the number of e-folds.
This is certainly  not be compatible with the quantum bound. We
should, however, note that at high energies in the early universe,
the quantum correction may be quite different than the one given by
Coleman-Weinberg potential and the simple analysis presented here
may not be valid in that regime.
\section{Conclusion}
In this letter, we have investigated the issues of classicality of
scalarons in the Starobinsky of $f(R)$ gravity model.{The model is
studied in the Einstein frame where we have General Relativity with
a scalar field, the scalaron. At the densities studied, local
analysis, we have shown that the curvature scalar is equivalent to
the one in General Relativity. The scalar field appears as a
Klein-Gordon massive field and can be quantized using the standard
procedure. In this context,} we have shown that the quantum bound on
scalaron mass derived in \cite{Upadhye:2012vh} can be a tight
constraint on $f(R)$ dark energy models. Within the range of
viability of the parameters of the model, that we derived, the
quantum corrections are large for low densities. The scalaron masses
increases very fast with medium density than the quantum bound on
it. The mass of the scalaron is unbounded and can exceed the Planck
mass at a reasonable value of the density of the medium. Clearly,
the model cannot be trusted in this case. We therefore need to
introduce a cut off to the mass of scalaron as the expression of our
ignorance and use the model below the cut off.\\
In view of the aforesaid, we used the extended Starobinsky model by
adding quadratic curvature correction to the Lagrangian.{We have
shown that this term does not change the curvature scalar $R \simeq
\rho/M_{pl}^2$ but only effects the form of the potential $V$ and
therefore the mass of the scalaron. This term influences principally
the scalar sector of the theory.} It produces a natural bound on the
mass of the scalaron that we constrained via the classicality
condition and the E\"ot-Wash experiments. We have demonstrated that
extended scenario is consistent with the quantum bound on the
scalaron mass.

{It  should also be noticed that in the regime of high densities or
low densities but large scales,  gravity is never weak and
quantization of the scalaron might become complicated. As noticed by
Starobinsky several years back, the quantum corrections in his model
during inflation are small\cite{staro1} and hence there is no strong
bound on the mass of scalaron from the classicality condition in
this case. The analysis performed in this letter is done in a regime
where the gravity is weak, density is low and scales are small.
Hence the standard results of quantum field theory could be
applied}. { Last but not least, we should clearly point out a subtle
point of our analysis. To be fare, underlying argument regarding the
quantum bound relies on the assumption of a semiclassical gravity in
which chameleon field is quantum and gravitational field classical.
But scalaron field has geometric origin which controls the curvature
of space-time. To be systematic, a full analysis should be
performed. Thus the scalaron field theory, which is obviously not
renormalizable, can be judiciously  used below some ultraviolet cut
off $M_{uv}$. The quantum corrected effective Lagrangian contains a
term proportional to $m^2_{eff} M^2_{uv}$ which might effect the
analysis presented here\cite{staro1}; in our opinion, the problem
requires further investigation.}

\section{ACKNOWLEDGEMENTS}
We are indebted to A. Starobinsky for taking pain in reading through
the draft and for making detailed comments on the theme of the
Letter in his kind communication. MS thanks S. Nojiri, S. Panda, K.
Bamba, R. Adhikari and Berry Christopher for useful discussions. MS
is supported by JSPS long term visiting fellowship and thanks
Kobayashi-Maskawa Institute for the Origin of Particles and the
Universe for hospitality. MS is also supported by the Department of
Science and Technology, India under project No. SR/S2/HEP-002/2008.
The work of RG is supported by the Grant-in-Aid for Scientific
Research Fund of the JSPS No. 10329.

\end{document}